\documentclass[submitted]{IEEEtran}
\usepackage{graphicx}
\usepackage{url}
\usepackage{cite}
\usepackage{color}
\usepackage{upgreek}
\usepackage{enumerate}
\usepackage[colorlinks=true,linkcolor=blue,urlcolor=blue,citecolor=blue]{hyperref}
\usepackage[colorinlistoftodos]{todonotes}
\usepackage{booktabs}
\usepackage{multirow}

\begin{document}

\title{\bf A Compact Magnet System for the Tsinghua Tabletop Kibble Balance}
\author{Yongchao Ma, Nanjia Li, Weibo Liu, Kang Ma, Wei Zhao, Songling Huang,\\ \textit{Senior Member, IEEE}, Shisong Li$^{\dagger}$, \textit{Senior Member, IEEE}

\thanks{The authors are with the Department of Electrical Engineering, Tsinghua University, Beijing 100084, China.
Wei Zhao is also with the Yangtze Delta Region
Institute of Tsinghua University, Jiaxing, Zhejiang 314006, China.}
\thanks{This work was supported by the National Key Research and Development Program of China under Grant 2022YFF0708600 and the National Natural Science Foundation of China under Grant 52377011.}
\thanks{$^\dagger$Email: shisongli@tsinghua.edu.cn}}

\maketitle

\begin{abstract}
Although the so-called magnetic geometrical factor, $Bl$, of a Kibble balance does not appear in the Kibble equations, it offers the precision link between electrical and mechanical quantities and furthers a quasi-quantum traceability path for mass metrology. This feature makes the magnet system, supplying the $Bl$ in Kibble equations, play a core role in Kibble balances. Following the open-hardware idea, we report here on the design, manufacture, assembly, optimization, and finally performance of a compact magnet system for the Tsinghua tabletop Kibble balance. Notably, the magnet system showcased in this study facilitates a straightforward upper levitation of splitting through a streamlined mechanism guide, substantially enhancing the ease of open and close operations.
Experimental tests show the realized magnet systems can yield a high $Bl$ value (e.g., 400\,Tm for a bifilar coil and 800\,Tm for a single coil with a wire gauge of 0.2\,mm) meanwhile a low volume/weight (40\,kg) thanks to the uniformity improvement of magnetic profiles. Furthermore, important parameters related to systematic effects, such as the current effect, are checked, aiming for a final mass-realization accuracy at the $10^{-8}$ level. 
\end{abstract}


\section{Introduction}

\IEEEPARstart{T}{he Kibble} balance, originally identified as the watt balance~\cite{Kibble1976}, stands as one of the principal methodologies for realizing the kilogram, the unit of mass, within the revised International System of Units (SI)\cite{cgpm2018}. An alternative approach, the x-ray crystal density (XRCD) method\cite{fujii2016realization}, represents the other major approach. Currently, numerous groups, predominantly national metrology institutes (NMIs), are engaged in Kibble balance experiments~\cite{NRC,NIST,NIST2,METAS,BIPM,LNE,MSL,NIM,KRISS,UME,PTB,NPL}, and the most precise Kibble balances exhibit the capability to calibrate masses at the kilogram level with a relative uncertainty of approximately one part in $10^8$~\cite{NRC,NIST}.

The detailed principle of the Kibble balance experiment has been summarized in recent reviews, e.g., \cite{Stephan16}. The principle is a combination of two simple physical laws, Lorentz's force law and Faraday's induction law, on the same magnet-coil system. In the so-called weighing phase, a current-carrying coil is placed in the magnetic field and the electromagnetic force is adjusted to balance the weight of a test mass, yielding $mg=(Bl)_\mathrm{w}I$, where $m$ is the test mass, $g$ is the local gravitational acceleration, and $I$ is the current in the coil. The $Bl$ is a geometric factor that integrates the cross product of the magnetic flux density $\vec{B}$ and the unit length vector along the wire $\vec{dl}$ over the entire coil wire path.  In the second measurement phase, the velocity phase, the coil is moved at a velocity $v$ and the induced voltage $U=(Bl)_\mathrm{v}v$ is obtained. Ensuring that $(Bl)_\mathrm{w}=(Bl)_\mathrm{v}$, the mass can be realized in terms of $m=UI/(gv)$. Since the quantities on the right-hand side can be measured against quantum standards, either electrical or optical, the Kibble balance can be considered a quasi-quantum mass realization instrument, see \cite{haddad2016bridging}. 

As can be seen from the measurement principle, although $Bl$ does not appear in the Kibble equations, it is a hidden key player: $Bl$ provides the precision link between two measurement phases so that the mass $m$ can be precisely determined. A detailed discussion of the irony of $Bl$ can be found in \cite{li2022irony}. It is shown in \cite{schlamminger2013design} that the value of $Bl$ should be neither too large nor too small. A large $Bl$ introduces uncertainty due to the small current measurement when the test mass is fixed, while the opposite, i.e. a small $Bl$ value, leads to high measurement uncertainty for a low induced voltage. It is found that a typical optimum $Bl$ value is a few hundred Tm for the kilogram-level mass realization. In order to maintain a suitable $Bl$ value while suppressing related systematic effects \cite{li18,linonlinear,linonlinear2,hysteresis,li2022} to a level well below $1\times10^{-8}$, a large field uniform region is desired. In this regard, the magnet system providing the magnetic field is typically large and heavy. {Nowadays, nearly all Kibble balance experiments worldwide utilize magnetic circuits composed of rare-earth permanent magnets and yokes \cite{li2022irony}.} To build such large magnet systems, the cost of the enormous magnet system is extremely high, and its assembly becomes difficult since the attractive forces for gluing the permanent segments as a whole or merging the permanent magnet and yoke parts can easily reach a few or a few tens of kN. 
In recent years, several groups have started the design of the compact magnet system, e.g.~\cite{NPL3,PTB,chao2020performance}, for tabletop Kibble balance applications. However, the compact magnet system may increase the magnetic uncertainties related to the hysteresis effect, the thermal effect, and the current effect\cite{li2022}. Therefore, it remains a challenging task to limit each related systematic effect while realizing a compact and easy-to-use Kibble balance magnet system. 

In late 2022, Tsinghua University launched a tabletop Kibble balance project for compact, robust, and accurate mass realizations following an open hardware approach~\cite{li2022design,li2023design}. In this paper, we report the design, fabrication, assembly, optimization, and performance of a compact magnet system for the Tsinghua Tabletop Kibble balance. Section \ref{Sec02} presents the design and general considerations of the magnet system, along with measurements of the magnetic material properties. Section \ref{Sec03} presents the machine and the assembly of various mechanical segments, in which we show that an appropriate splitting surface can significantly reduce the difficulty of the open/close operation of the magnet. Section \ref{Sec04} measures the magnetic profile, and some related systematic effects are discussed. Finally, a conclusion is drawn and the future work is discussed in Section \ref{Sec05}.

\section{General Design and Considerations}
\label{Sec02}

\begin{figure}[tp!]
    \centering
    \includegraphics[width=0.44\textwidth]{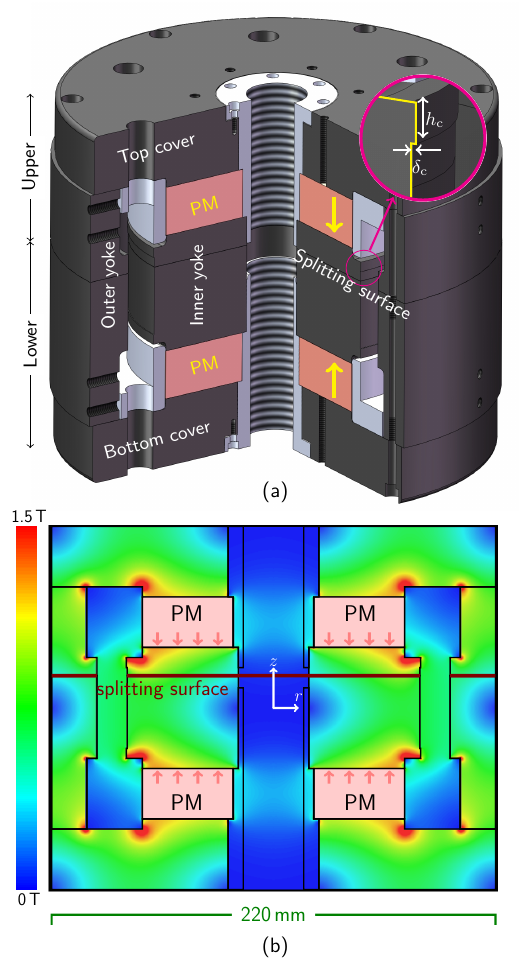}
    \caption{Design of the magnet system in the Tsinghua tabletop Kibble balance. (a) and (b) respectively show the CAD and schematic models. The background presented in (b) is the magnetic flux density map of the NdFeB-version permanent magnet system.}
    \label{fig:3d}
\end{figure}

\subsection{Basic Design}
Compared to other types of magnet systems, for example, current-carrying coils \cite{nimx,nist3}, the permanent magnet system can offer a sub-Tesla magnetic field without any current stabilization and ohmic heating issues. At present, almost all the world's Kibble balance experiments choose to use magnetic circuits with permanent magnets and iron yokes. Among different Kibble balance magnetic circuit designs, the one originally proposed by the BIPM Kibble balance group~\cite{BIPMmag2006} and later adopted by many other groups, e.g. \cite{NISTmag,METAS,KRISS,NIM,UME}, becomes the most popular. The magnet system design for the Tsinghua tabletop Kibble balance is also based on the BIPM-type magnetic circuit. The CAD model and schematic design are presented in Fig.  \ref{fig:3d} (a) and (b). The targeted magnet system has a 220\,mm outer diameter, 180\,mm height, and about 40\,kg in total mass. The flux of two permanent magnet rings, each denoted as PM, is guided by the inner yokes through a circular air gap formed by inner and outer yokes. The outer yoke and cover yokes return the flux as a closure. Note that in this design, we choose two different rare-earth materials, Neodymium (NdFeB) and Samarium Cobalt (Sm$_{2}$Co$_{17}$), and build two versions of magnet systems. In the following text, they are noted as respectively the NdFeB magnet and the SmCo magnet. For the yoke, a high permeability pure iron, DT4C, is chosen to form good equal potential boundaries at the inner and outer radii of the air gap to restrict the direction of the magnetic lines. The idea is to lower the influences of upper and lower PM asymmetries and ensure the generated magnetic field can follow a $1/r$ distribution inside the air gap~\cite{BIPMmag2017}, detailed in subsection \ref{yoke}. 

For the BIPM-type magnet circuit, the approximate formula to estimate the average magnetic flux density in the air gap, $B$, can be written as ~\cite{you2016designing,diamagnetic2020}
\begin{equation}
    B \approx -\mu_{0}H_{\rm m}/\left(\frac{\delta_{\rm a}}{\delta_{\rm m}}+\gamma\frac{S_{\rm a}}{S_{\rm m}}\right),
    \label{eq:01}
\end{equation}
where $\mu_{0}$, $H_{\rm m}$, $\delta_{\rm a}$, $\delta_{\rm m}$, $S_{\rm a}$, $S_{\rm m}$ are the vacuum permeability, the residual coercive force of the permanent magnet, the air-gap width, the permanent magnet height, {half of the air gap area along the flux path ($S_{\rm a}\approx \pi r_\mathrm{a}h_\mathrm{a}$ where $r_\mathrm{a}$, $h_\mathrm{a}$ are the mean radius and the height of the air gap) and the top/bottom surface area of the permanent magnet disk}, respectively. Here $\gamma$ is a scale factor due to the edge effect, mainly related to the height-width ratio of the air gap, i.e. $h_{\rm a}/\delta_{\rm a}$. In our design, the mean radius, width, and height of the air gap are respectively $r_\mathrm{a}=80$\,mm, $\delta_\mathrm{a}=15$\,mm, and $h_\mathrm{a}=50$\,mm. For such a geometrical setup, $\gamma\approx 1.43$ is obtained by the finite element analysis (FEA). The inner and outer radii of the permanent magnet ring are $r_{\rm mi}=20$\,mm and $r_{\rm mo}=65$\,mm, and its height $\delta_{\rm m}=25$\,mm. $H_{\rm m}({\rm NdFeB})=-1030$\,kA/m and $H_{\rm m}({\rm SmCo})=-786$\,kA/m are obtained by FEA simulations based on the surface magnetic flux density measurement, see subsection \ref{pm}.  Taking the above parameters into (\ref{eq:01}), the average magnetic flux density is 0.46\,T and 0.62\,T respectively for SmCo and NdFeB magnets.

\begin{figure}[tp!]
    \centering
    \includegraphics[width=0.475\textwidth]{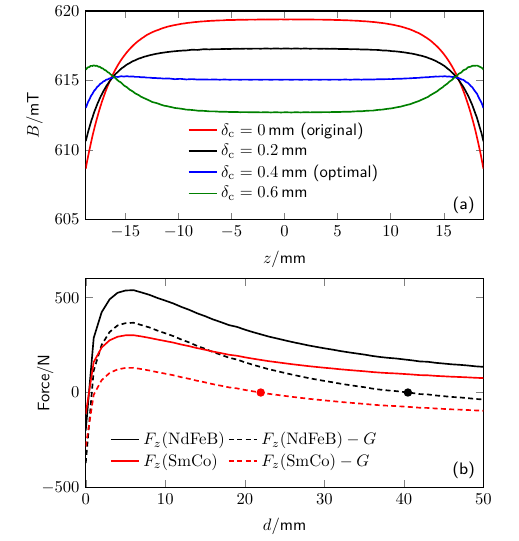}
    \caption{(a) presents the magnetic profiles with different inner yoke shape compensations. In the shown case, the height of the rectangle is fixed at $h_{\rm c}=5$\,mm and its width $\delta_{\rm c}$ varies from 0\,mm (original) to 0.6\,mm. The optimal $\delta_{\rm c}=0.4$\,mm is employed in the Tsinghua tabletop Kibble balance. (b) shows the magnetic attraction force as a function of the splitting distance, $d$. The minus sign means the force is attractive while the positive sign denotes a repulsive force. The red and black dots are the levitation locations of the upper segment for respectively the SmCo and NdFeB systems.}
    \label{fig:feature}
\end{figure}

During the Kibble balance measurements, uniformity of the magnetic field at the centimeter {range} along the vertical direction, e.g. $\Delta B/B<1\times10^{-3}$, is desired to ensure the uncertainty of the measurement results. However, due to the edge effect, the magnetic flux density of a conventional BIPM-type magnet drops significantly {{at two ends}} of the air gap, leaving a much shorter field uniform range than the height of the wide air gap. Fig. \ref{fig:feature} (a) shows the magnetic profile of the conventional BIPM-type magnet using the same air-gap parameters. The uniform range ($\Delta B/B<1\times10^{-3}$) is about 23\,mm, approximately 46\% of the whole air gap height. Since for tabletop Kibble balances, the volume of the magnet system is limited, it is crucially necessary to extend the uniform field range without enlarging the overall magnet size. 

An innovation of the Tsinghua tabletop Kibble balance magnet lies in the optimization of its profile through a strategic reshaping of the inner yoke boundary. This approach draws inspiration from the magnet design proposed in \cite{ss20}, incorporating two small rectangle rings at the extremities of the inner yoke to mitigate the edge effect. The rationale behind this modification is elucidated by (\ref{eq:01}): The reduction of the air gap width ($\delta_{\rm a}$) at both ends effectively amplifies the magnetic field $B$. Fine-tuning the dimensions, specifically the width ($\delta_{\rm c}$) and height ($h_{\rm c}$) of the rectangle, enables significant compensation for the edge effect, thereby expanding the range of uniform magnetic fields.
Fig. \ref{fig:feature}(a) illustrates our approach, where we maintain a fixed height for the compensation rectangle ring ($h_{\rm c}=5$ mm) and vary its width ($\delta_{\rm c}$) from 0\,mm to 0.6\,mm through FEA simulations. It is evident that excessively small widths result in insignificant compensation, while excessively large widths lead to an over-compensation, reducing the uniform range. In the final design, a width of 0.4\,mm is chosen for the compensation ring. Despite a modest decrease of approximately 4.3\,mT (0.7\%) in the absolute magnetic field strength, this configuration enhances the uniform magnetic field range ($\Delta B/B<1\times10^{-3}$) to 35\,mm, representing 70\% of the total air-gap height. Notably, this signifies a more than 50\% increase in the uniform range compared to the original uncompensated magnet system. The experimental validation of the magnetic profiles is discussed in Section \ref{Sec04}.

The second major improvement of the Tsinghua Kibble balance magnet, compared to the conventional BIPM-type design, is the optimization of the open/close operation. The BIPM-type permanent magnet contains a magnetic closure that can well {reject} magnetic interactions from the background flux, however, a drawback of such systems is that a high attraction force (up to the ton level and above) needs to be overcome for magnet open/close operations so that the coil can be reached (install, replace, adjustment, etc). It is obvious that the smaller the force required to open the magnet system, the easier it is to maintain the magnet system. 

The ideal case is to choose an operation surface that has the minimum change to the main magnetic flux path between the open and the close status. With a symmetrical design, the middle horizontal plane ($z=0$) seems to be an option because all the magnetic flux on this surface is horizontal, i.e. $B_z=0$. However, FEA calculations \cite{li2022irony} and theoretical analysis using Maxwell's tensor \cite{marangoni2019magnet} show that the splitting force on the middle symmetrical surface is repulsive. This phenomenon is understandable: Once the two symmetrical parts are separated, the edge effects produce the same magnetic poles around the air gap region, and a repulsive force is hence generated. 

In \cite{marangoni2019magnet},     
the approximation of the vertical splitting force, $F_z$, as a function of the splitting position $z$ is given, i.e.
\begin{equation}
    F_z (z) = \frac{\mathrm{\pi}r_\mathrm{a} B^2}{\mu_0} \left(\delta_\mathrm{a} -\frac{2\mathrm{\pi}r_\mathrm{a}(S_{\rm{i}}+S_{\rm{o}})}{S_{\rm{i}}S_{\rm{o}}} z^2\right),
    \label{eq:02}
\end{equation}
where $r_\mathrm{a} = (r_\mathrm{i}+r_\mathrm{o})/2$ denotes the average radius of the air gap, $S_{\rm{i}}$, $S_{\rm{o}}$ are respectively the cross-section area of the outer and inner yokes at the splitting plane, $B$ is the average magnetic flux density in the air gap, and the $z$ means the distance of the split plane to the symmetry plane of the magnet. Although (\ref{eq:02}) estimates only the magnetic force with an opening distance $d=0$\,mm, it clearly shows how the $F_z$ varies from repulsive to attractive. Also, the magnetic force $F_z$ varies as a function of the opening distance, $d$ and the FEA mapping of the magnetic force $F_z(z,d$) \cite{li2022irony} shows that the sign of $F_z$ can also revise, from attractive to repulsive, at the $z$ regions that $F_z(z,d=0\,\mathrm{mm})<0$. Summarizing the above information, it is possible to find a $z$ value as the open/close surface and ensure the overall magnetic force $F_z$ minimum. In addition, considering the splitting surface may change the shape of the profile, it requires pushing as far away from the middle symmetry as possible. In our design, the splitting position $z$ is set as 15\,mm. Fig. \ref{fig:feature}(b) presents the magnetic force as a function of the splitting distance, $d$, in a range of 50\,mm. The initial opening force ($d=0$\,mm) is -200\,N and -112\,N respectively for the NdFeB and SmCo magnets. Within a few millimeters, $F_z$ changes direction from attractive to repulsive and then quickly reaches the maximum (approximately 540\,N and 302\,N at $d=6$\,mm). Then finally the force $F_z$ goes down slowly when $d$ continues to increase. 

In Fig. \ref{fig:3d}, the magnet system is delineated by a splitting surface, dividing it into distinct upper and lower segments. The respective masses of these segments are 17.2\,kg and 24.1\,kg. In the design, the lower half is immovably fixed, while the upper half remains mobile. Fig. \ref{fig:feature}(b) illustrates that subtracting the weight of the upper half, $G=168$\,N, from $F_z$ results in a zero-crossing point denoted by a prominent dot in the plot. At this specific point, $F_z$ equals $G$, leading to the levitation of the upper half magnet. This characteristic stands as a distinctive feature of the Tsinghua Kibble balance magnet system. Notably, the considerable gap during levitation measures 22\,mm and 41\,mm for the SmCo and NdFeB magnets, respectively. This ample gap accommodates coil operations effectively. {{It is important to emphasize that in the closed position, the attractive forces are 280\,N and 368\,N for SmCo and NdFeB magnets, respectively.}} This configuration ensures a securely closed state while maintaining a reasonable ease of reopening.

\subsection{Material Properties}

\subsubsection{Permanent Magnet}
\label{pm}
The permanent magnet is the magnetic source of the magnet system. Its properties are closely related to the stability and uniformity of the air-gap magnetic field. Here we compare two types of permanent magnets: Samarium-cobalt (Sm$_2$Co$_{17}$) magnet and Neodymium (NdFeB) magnet. Since the temperature coefficient of NdFeB magnet ($\approx-1.2\times 10^{-3}$/K) is a few {times} of the SmCo magnet ($\approx-3.0\times 10^{-4}$/K), most Kibble balance magnets choose SmCo as the permanent magnet material.
{
Due to the brittleness of SmCo material, there is a size limitation in its manufacturing process. As a result, large SmCo magnets are typically constructed by assembling multiple smaller magnet tiles. However, variations in the magnetization of these tiles, along with the gaps introduced during assembly, can significantly degrade the overall uniformity of the magnetic field in the permanent magnet.} For the required size of the permanent ring ($r_{\rm mi}=20$\,mm, $r_{\rm mo}=65$\,mm, $\delta_{\rm m}=25$\,mm), it is still too large for a whole {magnetization}. In the design, we divided the permanent magnet ring into 8 slices. Conventionally, each slice is magnetized first (choose slices with a similar magnetization, see \cite{NISTmag}) and then all slices are glued into a whole. The latter needs to overcome the attraction force between adjacent slices and the assembly is usually costly. Here we use a different sequence for the magnetization: the slices are first assembled (glue or use an outer ring) and then individual slices are magnetized one by one. This approach can significantly reduce the cost, but the drawback is that the magnetization of each slice may differ considerably.  

\begin{figure}[h]
    \centering
    \includegraphics[width=0.475\textwidth]{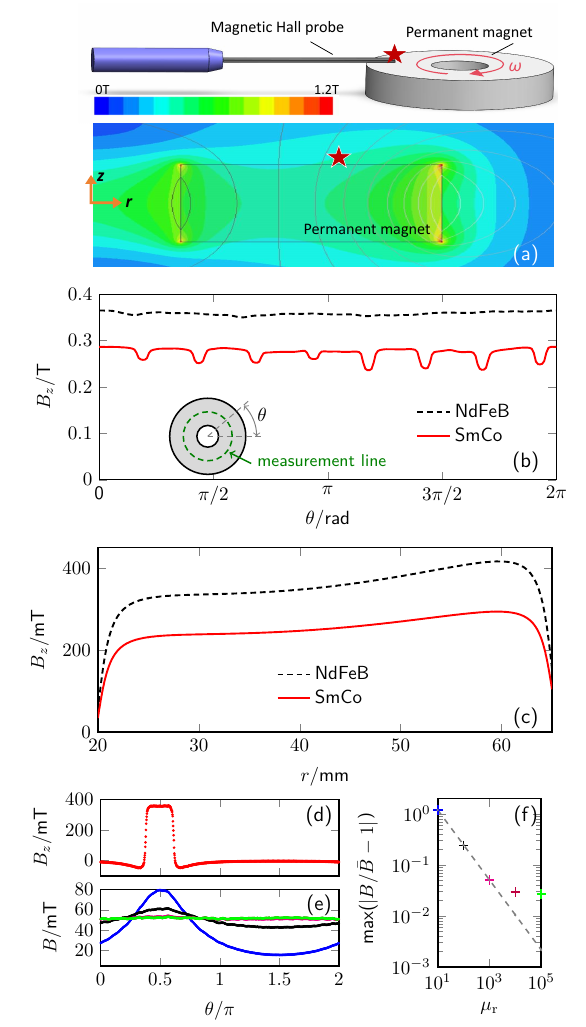}
    \caption{(a) shows the surface magnetic field measurement setup. The lower plot is a $B$ field map around the permanent disk (NdFeB). (b) shows a typical measurement result of the magnetic flux density on the top surface of permanent magnet rings. The measurement was carried out along the middle circular line (green dashed line) and $\theta$ is defined as the angle of the scan, from 0 to $2\pi$ shown in the subplot. (c) shows the vertical magnetic field, $B_z$, as a function of $r$ for two versions of permanent magnets. (d) presents the surface magnetic field distribution of the permanent magnet used for 3D FEA simulation. Only one-eighth of the permanent magnet disk ($\theta=\pi/2$) is set magnetized. (e) shows the magnetic field uniformity in the air gap center with $\mu_{\rm r}=$10, 100, 1000, 10000, and 100000 respectively. (e) presents the maximum variation of the $B$ field in (e) as a function of $\mu_{\rm r}$. The dashed line is the theoretical value and the points are obtained using FEA. Different colors are related to different $\mu_{\rm r}$ values as presented in (e). }
    \label{fig:PM}
\end{figure}

To check the uniformity of different slices, a measurement of the surface magnetic field was carried out. The experimental setup is shown in Fig. \ref{fig:PM}(a). A Hall probe is placed slightly above the middle surface ($r=45$\,mm) and the vertical magnetic field is scanned along 360 degrees when the permanent magnet disk is rotated by a motorized stage. One typical measurement result is shown in Fig. \ref{fig:PM}(b) (the red curve). It can be seen from the measurement that the magnetic field at the junction of two permanent magnets is significantly weaker (up to 15\%) than that in the middle of the permanent magnet slice, leaving 8 magnetic poles along the $\theta$ axis. Using the same experimental setup, the surface magnetic field of the NdFeB ring is also measured, shown as the black dashed line. Note the NdFeB disk is manufactured and magnetized as a whole as its magnetization size limit is much larger than that of the SmCo magnet. It can be seen that the curve of the NdFeB magnet is much smoother and the peak-peak variation is at the percentage level. Also, the magnetic field generated by the NdFeB magnet is about stronger, and the surface magnetic flux density increases by over 0.1\,T, from 0.25\,T to 0.36\,T. 

Using the surface field measurement results, the residual coercive force of the permanent magnet, $H_{\rm m}$, can be conducted. The idea is to adjust and find the right $H_{\rm m}$ in an open space FEA model so that the vertical magnetic field at the Hall probe measurement point equals to the measurement result. As shown in Fig. \ref{fig:PM}(c), the final $B_z$ distribution along $r$ is presented, and the $H_{\rm m}$ values for the test NdFeB and SmCo magnets are $-1030$\,kA/m and $-786$\,kA/m, respectively. 

Knowing the main parameters of permanent magnet rings, a concerning question is how well can the yoke average out the asymmetry of the permanent magnet rings, such as the ripples shown in Fig. \ref{fig:PM}(b). The asymmetry of the upper and lower permanent magnets can be well rejected by using two permanent magnets with a similar surface field strength and high permeability yokes, see \cite{BIPMmag2017}. For example, for the NdFeB magnet, the variations of the surface field is within 4\%. In this case, according to the evaluation in \cite{BIPMmag2017}, even with a relatively low permeability yoke, e.g. $\mu_{\rm r}=1000$ for low-carbon steel, the relative magnetic field change over $z$ is well below $1\times10^{-3}$. By far, the field uniformity along the circular, $B(\theta)$, has not been studied. To observe the magnetic field uniformity change as a function of yoke permeability, a 3D FEA model is employed. For the permanent magnet ring, only one-eighth located at $\theta=\pi/2$ is magnetized and used as the flux input. Its surface magnetic field distribution is shown in Fig.   \ref{fig:PM}(d). To keep the setup {simple}, the upper and lower permanent magnet rings are kept the same field distribution along $\theta$. Now we change the relative permeability of the yoke, from 10 to $10^5$, and calculate the field distribution in the air gap. The FEA calculation results are shown in Fig.  \ref{fig:PM}(e). It can be seen that as the $\mu_{\rm r}$ increases the field becomes flatter and flatter along the $\theta$. Limited by the meshing numbers, the calculation starts to {lose} accuracy when $\mu_{\rm r}$ is larger than 1000. 
But we can extrapolate the calculation results and find out the $B$ uniformity improvement as a function of $\mu_{\rm r}$. As shown in Fig.  \ref{fig:PM}(f), the variation of the $B(\theta)$ profile, defined as ${\rm max}(|B/\bar{B}-1|)$ where $\bar{B}$ is the average of $B(\theta)$ over 0 to $2\pi$, as a function of the relative permeability of the yoke is presented. The cross points are values calculated by FEA from Fig.  \ref{fig:PM}(e) and the dashed line is an {extrapolation} using the first three points, where the FEA calculation has not yet been limited by the meshing. It can be seen that with $\mu_{\rm r}>10^4$, the uniformity of $B(\theta)$ can be improved over two magnitudes. For the SmCo rings shown in Fig. \ref{fig:PM}(b), if the eight peaks or valleys differ at the percentage level and the yoke permeability is at $10^4$ level, the symmetry of $B(\theta)$ can be suppressed to the $10^{-4}$ level, which is largely enough for the coil alignment adjustment. For NdFeB magnet, since its surface flux density change of the permanent disk is much less along $\theta$, the asymmetry of the permanent magnet disk is not a limiting factor for obtaining a uniform $B(\theta)$ profile.  

\subsubsection{Magnetic Yoke}
\label{yoke}
As discussed in the above section, high-permeability yokes are extremely important for eliminating permanent asymmetries and ensuring the uniformity of the magnetic field in the air gap. By far, there are three categories of yoke materials employed in constructing the Kibble balance magnet: low-carbon steel, pure iron, and iron-nickel alloy (e.g. Supra50). Their relative permeability $\mu_{\rm r}$ is respectively at the $10^3$, $10^4$, and $10^5$ order, and the saturation field is about 1.8\,T, 1.5\,T and 1.2\,T~\cite{li2022irony}. For tabletop systems, a high saturation field can help to reduce the yoke volume and hence the overall size of the magnet. A comprehensive consideration of two factors, the Tsinghua Kibble balance magnet uses pure iron, DT4C, as the yoke material. The carbon fraction of DT4C usually is below 0.025\%. Heat treatments can recover the permeability loss during the machining process, and the DT4C requires a deoxygenation annealing in {a vacuum or hydrogen atmosphere}. Here we measured the magnetic permeability of the DT4C sample before and after annealing in vacuum.

\begin{figure}
    \centering
    \includegraphics[width=0.45\textwidth]{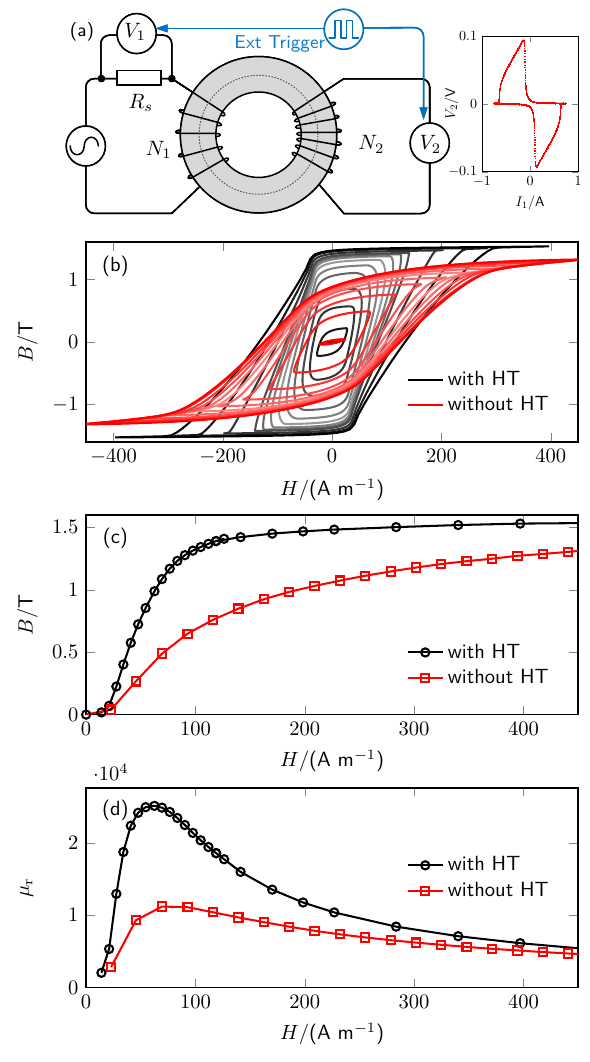}
    \caption{(a) presents the experimental setup for measuring the yoke permeability. {The right subplot is an example of excitation current in the primary winding $I_1$ and the induced voltage in the secondary winding $V_2$.} (b), (c) and (d) respectively show the hysteresis curve, the $BH$ curve, and the $\mu H$ curve of two yoke samples. HT denotes heat treatment. }
    \label{fig:yoke}
\end{figure}

The $BH$ curve and magnetic permeability of the samples are measured by electromagnetic induction. The measurement principle diagram is shown in Fig. \ref{fig:yoke}(a). The samples are machined into the rings and two sets of coils, the excitation coil $N_1=480$ {turns} and induction coil $N_2=280$ {turns}, are {wound} on the ring. The primary coil is excited by the output of a signal generator with a low frequency, 0.1\,Hz. The current through the primary is measured by a 10\,$\Omega$ sampling resistance, $R_s$. The voltage drop of $R_s$ and the induced voltage of the secondary coil are synchronized and measured by two digital voltage meters (3458A, $V_1$ and $V_2$). The $H$ field through the yoke ring is obtained by Ampere's law, i.e.
\begin{equation}
    H=\frac{N_{1}V_1}{lR_s},
\end{equation}
where $l=2\pi r_0$ ($r_0=25$\,mm is the mean radius of the ring). The induced voltage on the secondary coil is written as
\begin{equation}
    V_2=N_2 A \frac{{\rm d}B}{{\rm d}t},
\end{equation}
where $A=100$\,mm$^2$ is the cross-section area of the iron ring.  The $B$ field then can be calculated  as
\begin{equation}
    B=\int_T\frac{V_2}{N_2 A}{\rm d}t-B_0.
\end{equation}
Note that $B_0$ is a constant that makes the average of the $B$ field in a period  $T$ equal to zero. 
 
 The measurement results, including the hysteresis curve, the $BH$ curve, and the $\mu H$ curve, are shown in Fig.  \ref{fig:yoke}(b)-(d). As expected, a significant increase of permeability is observed after the heat treatment.  {The results of the FEA simulation show that along the main flux path, the magnetic field in the yoke varies from about 40\,A/m to 200\,A/m.} Under such working status, the relative permeability $\mu_{\rm r}$ of the yoke is over $2\times 10^4$, which can well meet the requirement for realizing the designed target. One additional merit of using a high-permeability yoke is to achieve good magnetic shielding performance, offering a reduction of electromagnetic noise, as well as some systematic effects.

\section{Machine and Assembly}
\label{Sec03}

The substantial attractive force between the yoke and the permanent magnet renders the assembly of the magnet system a demanding task. Typically, the installation of the yoke and permanent components necessitates specialized tools or machinery, as referenced in \cite{NISTmag}. In this paper, we outline a straightforward procedure for assembling the designed magnet system using widely available tools. We detail the attraction force between various components and provide a step-by-step guide for installing each segment. Our approach is grounded in the open-hardware philosophy, enabling experimenters to replicate the magnet system by following the provided assembly instructions.

\subsection{Assembly Process}
\label{Subsec: Assembly}

\begin{figure}
    \centering
    \includegraphics[width=0.4\textwidth]{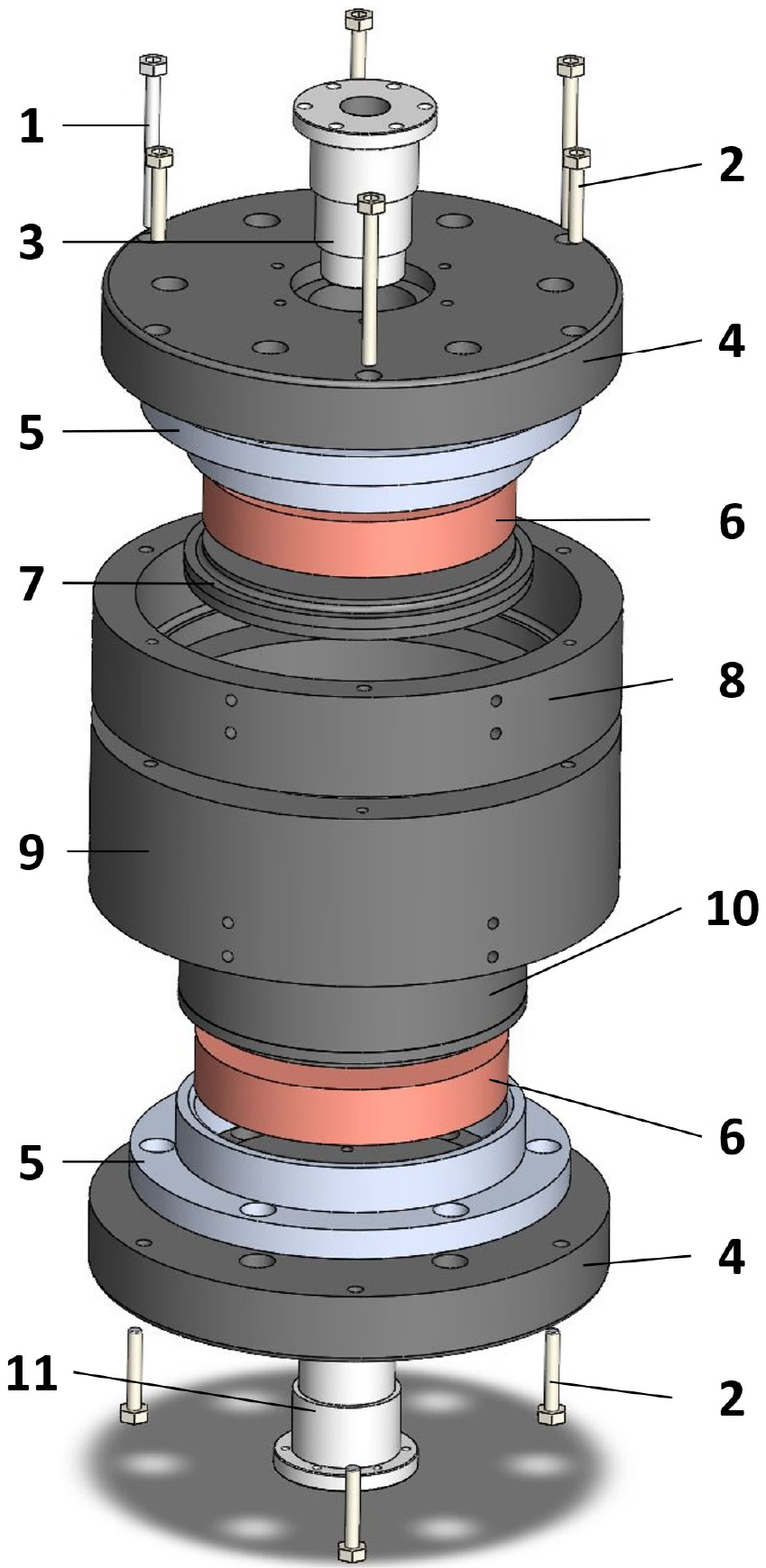}
    \caption{Overview of segments in the Tsinghua tabletop Kibble balance magnet system. 1-screws to secure the upper and lower halves. 2-screws to lock the cover yoke and the outer yoke. 3-aluminum bar for centering the upper cover yoke, upper permanent magnet and the upper inner yoke. 4-cover yoke. 5-aluminum guide ring. 6-permanent magnet. 7-inner yoke (upper). 8-outer yoke (upper). 9-outer yoke (lower). 10-inner yoke (lower). {11-aluminum bar for centering the lower cover yoke, lower permanent magnet and the lower inner yoke.}}
    \label{fig:seg}
\end{figure}

\begin{figure*}
    \centering
    \includegraphics[width=\textwidth]{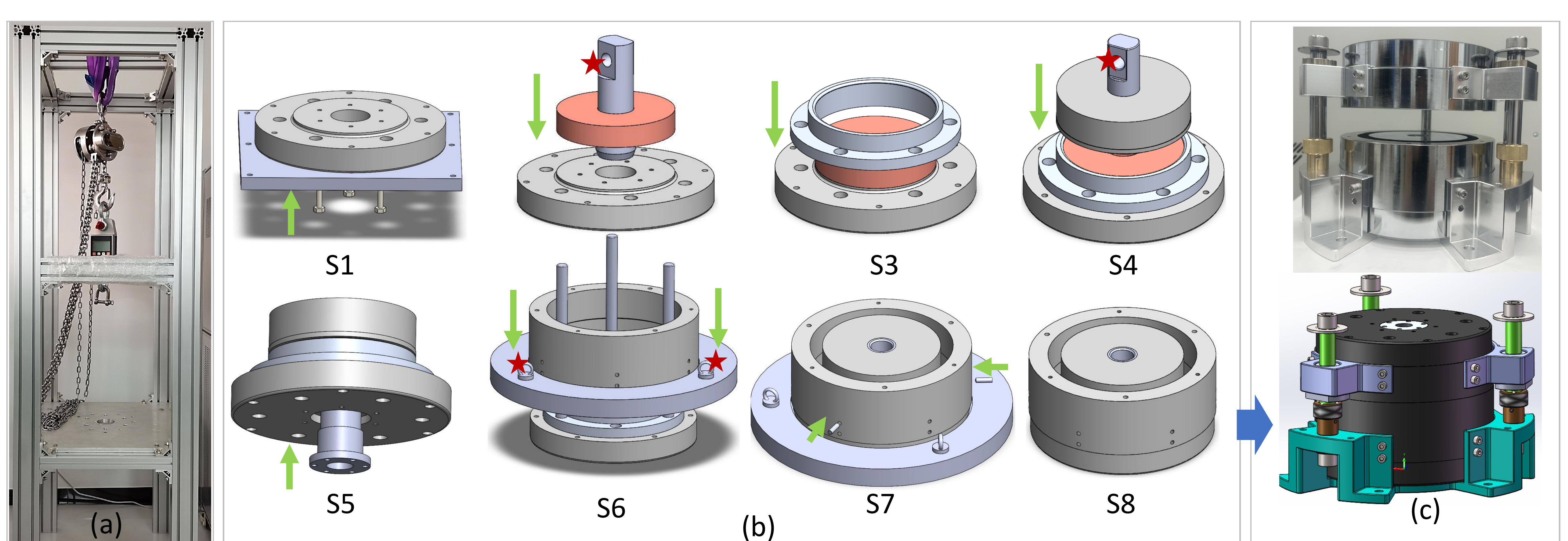}
    \caption{The assembly process of the magnet system. (a) shows the assembly cage. (b) presents the steps to install the lower half of the magnet. The upper half can be similarly assembled. The arrows denote the movement direction and the stars are the holding positions of the lifting rods connected to the assembly cage. (c) shows a mechanical mechanism to open/close the upper and lower parts of the magnet. The upper {illustration} is a picture of the magnetic levitation of the upper half magnet. }
    \label{fig:assemble}
\end{figure*}

As previously noted, the magnet system is divided into upper and lower sections to facilitate coil maintenance. Fig. \ref{fig:seg} illustrates the different components contained within these two halves. Generally, the assembly process follows a bottom-to-top and inside-to-outside sequence. Initially, the upper and lower parts of the magnet are assembled separately. Subsequently, these two sections are combined. Given that the assembly processes for the lower and upper parts are identical, we will use the lower part as an example to detail the steps. As depicted in Fig. \ref{fig:assemble}(b), the assembly procedure of each half is divided into 8 steps (S1-S8), listed as follows:

\begin{enumerate}[S1 -]
    \item \textit{Secure the cover yoke to the assembly platform.} Utilize the six threading holes of the cover yoke to mount it on the bottom plate of the assembly cage shown in Fig.   \ref{fig:assemble}(a). The design of the assembly cage and its accessories is discussed in subsection \ref{Subsec: Force_analysis}.
    
    \item \textit{Installation of the permanent magnet.} In this step, the permanent magnet is placed on a stainless steel bar. The diameter of the bar's end is larger than the inner hole of the permanent magnet but smaller than the diameter of the central hole of the cover yoke. This design allows the bar to support the magnet via the lifting rod in the assembly cage. After the permanent magnet is installed, the bar is removed from the bottom through the central hole of the yoke cover.

    \item \textit{Installation of the aluminum ring.} The aluminum ring serves to secure the permanent magnet and enhance its alignment. By aligning the six primary holes (intended for the coil rods and optics) of the aluminum ring with those of the yoke cover, the centering of the permanent magnet is improved. Furthermore, the aluminum ring plays a crucial role in adjusting the concentricity between the inner and outer yokes, as detailed in step S7.

    \item \textit{Installation of the inner yoke.} This step is similar to S2, with the primary difference being the diameter of the lifting bar. In this case, the bar is designed to support the inner yoke while also passing through the central hole of both the permanent magnet and the cover yoke.

    \item \textit{Installation of the aluminum centering bar.} The aluminum bar with four different diameter levels is inserted into the central hole of the inner yoke, the permanent magnet, and the cover yoke. Once the aluminum bar is fully inserted, its outer edge is secured to the bottom of the cover yoke with six mini screws.

    \item \textit{Installation of the outer yoke.} In the design, the outer diameter of the outer yoke is slightly larger than the outer diameter of the yoke cover. An aluminum ring, whose inner diameter is in between, is used to hold and guide the outer yoke from the suspension. It is important to insert three aluminum bars through the main holes for the isolation of the inner and outer yokes in case of a short circuit. When the outer yoke is approaching the cover yoke, a rough alignment of the screw holes should be carried out. 
    After the outer yoke touches the cover yoke, make sure three locking screws are in place but not tightened, and the three aluminum bars can be removed. 

    \item \textit{Concentric adjustment of the inner and outer yokes.} The screws locking the outer yoke and the cover yoke are kept loose, and insert three headless screws to the lower side threads every 120 degrees. By pushing the depth in different directions, the concentricity of the inner and outer yokes can be adjusted. The details of the adjustment are presented in subsection \ref{Subsec: Adjustment_concentricity}.

    \item \textit{Finalization of the outer yoke assembly.} Once the concentricity of the inner and outer yokes is optimized, the locking screws can be tightened. The cover yoke can be unlocked and the outer aluminum ring can be removed.
\end{enumerate}

The assembly and adjustment of the upper half of the magnet follow the same process as outlined above. Once both halves are installed, we proceed to the final step: combining the upper and lower parts. In this design, three long screws, labeled 1 in Fig.  \ref{fig:seg}, pass through the upper outer yoke and are threaded into the lower outer yoke. It is important to note that misalignment of the two halves can potentially short-circuit the magnetic path in the worst-case scenario, necessitating a guiding mechanism. Fig.  \ref{fig:assemble}(c) presents a mechanical design for opening and closing the upper and lower halves. Experimental tests validate the proposed concept of magnetic levitation for the upper part, as illustrated in the upper {illustration} showing the NdFeB system. The levitation is super stable with a separation distance of approximately 40\,mm, which offers enough space for operations such as inserting, removing, or modifying the coil. To close the magnet, a simple downward push with hand-scale force is sufficient. As mentioned, since the attraction force (including gravity) under closed status is 368\,N and 280\,N for NdFeB and SmCo magnets {respectively}, three nuts are attached to the guide bars, and by turning these nuts the magnet can be reopened. 

\subsection{Lifting Force and Device}
\label{Subsec: Force_analysis}

\begin{figure}[t!]
    \centering
    \includegraphics[width=0.475\textwidth]{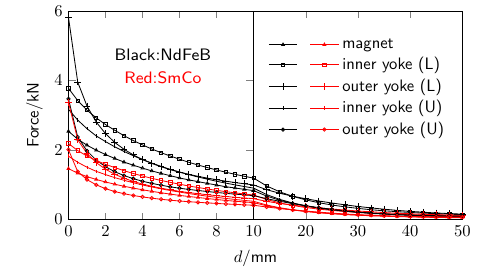}
    \caption{The magnetic attraction force at different assembly steps as a function of the separation distance. }
    \label{fig:Oforce}
\end{figure}

To counteract the substantial attractive force between the permanent magnets and the yokes, the use of a lifting device is required~\cite{NISTmag}. Selecting the appropriate capacity for the crane, along with a suitable weighing sensor, necessitates an evaluation of the maximum attractive force. The assembly force was analyzed by using FEA calculations. Since the yokes and permanent magnets move slowly during assembly, the impact of the eddy current effect is negligible. Consequently, it is appropriate to employ a static magnetic field model to analyze the force variation. The results, depicted in Fig.   \ref{fig:Oforce}, indicate that the attractive force is significant only when the distance between the yoke and the permanent magnet is small. The attraction force decays quickly with the separation distance and is well below 0.2\,kN at $d=50$\,mm. In the other cases, the gravitational force on the component is the dominant force. Tab. \ref{tab:Force} presents the maximum values of the force during the assembly steps considering the gravity of each segment connected to the lifting rods.

Based on the simulation results, we designed the lifting system, depicted in Fig. \ref{fig:assemble}(a). The system comprises a trestle, a retainer plate, a manual hoist, and metal hooks. To mitigate the attractive force generated by the permanent magnet, all components of the lifting system are constructed from non-ferromagnetic materials. The magnet system is secured to the retainer plate, which is centrally positioned on the trestle. This arrangement ensures that the attractive force between the yoke and the permanent magnet acts as an internal force within the lifting system, significantly reducing the tensile strength requirements for the trestle. The trestle is fabricated from 4080 aluminum profiles, with tensile and bending strengths exceeding 205\,MPa and 157\,MPa, respectively. The retainer plate is made from an aluminum plate, while the manual hoist and hooks are constructed from 304 stainless steel. During the assembly process, the maximum force encountered is approximately 5815\,N (equivalent to 593\,kg). Consequently, the selected manual hoist has a maximum lifting capacity of 1.5\,t.
To check these numbers, the lifting force is measured by a tension meter during the installation. The maximum force observed is much lower than the theoretical values in Tab. \ref{tab:Force}. This is because the maximum force occurs at $d=0$\,mm, and drops quickly once they are separated, and for the force measuring device, it is difficult to capture the opening or closed moment, $d=0$\,mm. Although the measurement result is much lower, the theoretical values shown in Tab. \ref{tab:Force} are still the most important references for designing and realizing the lifting system.   

 \begin{table}[tp!]
     \centering
     \caption{The maximum pulling force in each assembly step for both the SmCo and the NdFeB systems.}
     \begin{tabular}{l r r r r }
     \hline\hline
       \multirow{2}*{STEP} & \multicolumn{2}{c}{The lower part} & \multicolumn{2}{c}{The upper part}\\
        & SmCo\,/N & NdFeB\,/N & SmCo\,/N & NdFeB\,/N \\
       \midrule
        S1 & 82 & 82 & 82 & 82 \\
        S2 & 1466 & 2539 & 1466 & 2539 \\
        S3  & 6 & 6 & 6 & 6 \\
        S4 & 2188 & 3775 & 1826 & 3171 \\
        S5 & 3 & 3 & 2 & 2 \\
        S6 & 3365 & 5815 & 2010 & 3472\\
        S7 & --- & --- & --- & --- \\
        S8 & --- & --- & --- & --- \\
          \hline\hline
     \end{tabular}
     \label{tab:Force}
 \end{table}

\subsection{Concentricity Adjustment of Inner and Outer Yokes}
\label{Subsec: Adjustment_concentricity}
The target of a Kibble balance magnet system is to generate a one-dimensional radial magnetic field in the air gap: Ideally, the magnetic field should be uniform along the vertical and have a $1/r$ distribution along the radial direction. Since the magnetic flux density is inversely proportional to the gap width~\cite{li2022irony}, the concentricity of the inner and outer yokes can significantly affect the tangential uniformity of the radial magnetic field. Although, in theory, the decentering of inner and outer yokes does not affect the $Bl$ value \cite{li2016discussion} and can be compensated by the coil alignment, in practice, the gap between coil former and yokes is very limited, and the coil may touch the yoke before the alignment is done. In this case, the concentricity of the inner and outer yokes must be adjusted.

\begin{figure}
    \centering
    \includegraphics[width=0.45\textwidth]{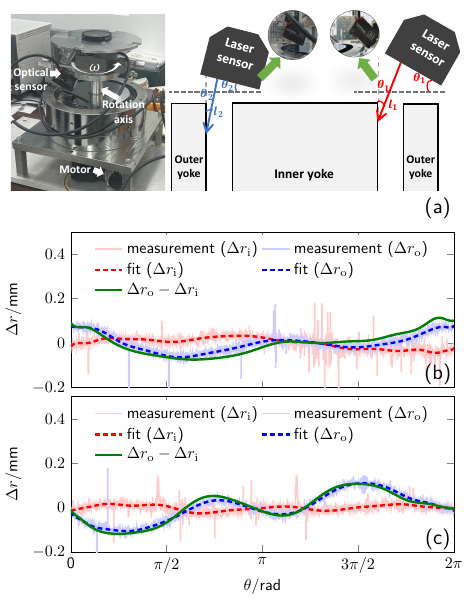}
    \caption{(a) shows the experimental setup for concentricity measurement of inner and outer magnetic yokes. (b) and (c) are the measurement results of the upper and lower part after adjustment, respectively.}
    \label{fig:8}
\end{figure}

Bielsa \textit{et al} proposed a concentricity optimization of inner and outer yokes using a double capacitor sensor~\cite{bielsa2015alignment}. However, in our case, the mean radius of the air gap is much smaller and the range of the capacitor sensor becomes limited when measuring the inner surface of the outer yoke. To address this issue, we use an optical sensor approach: As shown in Fig.~\ref{fig:8}(a), two laser sensors (Mirco-Epsilon 1420 with a measurement range 10\,mm, measurement repeatability 0.5\,$\upmu$m) are mounted in the same rotational stage. The measurement points are set respectively on the outer surface of the inner yoke and the inner surface of the outer yoke. After the stage is leveled and two sensors are adjusted in the measurement range, the variation of the gap width between the inner and outer yokes is given as  
\begin{equation}
    \Delta r = \Delta l_2\sin\theta_2-\Delta l_1\sin\theta_1,
\end{equation}
where $l_1$ and $l_2$ are distances measured by the two optical sensors shown in Fig.~\ref{fig:8}(a); $\Delta l_1$ and $\Delta l_2$ are residuals after removing their average values; $\theta_1$ and $\theta_2$ are the sensor mounted angles referred to the horizontal plane.  

The first step of the measurement is to make the centers of the inner yoke and the rotation stage coaxial. The inner yoke is mechanically fixed and the horizontal position of the rotation stage is adjusted to a status that the amplitude of the sinusoidal component with $2\pi$ period is minimum or comparable to its harmonics. Then fix the position of the rotation stage. Now, by rotating the stage with two optical sensors, the variations of the inner surface of the outer yoke, i.e. $\Delta r_\mathrm{o}=\Delta l_2\sin\theta_2$, and the outer surface of the inner yoke, $\Delta r_\mathrm{i}=\Delta l_1\sin\theta_1$, as a function of rotation angle $\theta$ can be obtained. Their difference, $\Delta r_\mathrm{o}-\Delta r_\mathrm{i}$, can define the concentricity of the inner and outer yokes. 
As presented in the assembly process S7, the screws locking the yoke cover and the outer yoke remain loose, and adjusting the side screws can push the aluminum ring and hence the position of the outer yoke in the backward direction. Keep the adjustment and measurement interactions until $\Delta r_\mathrm{o}-\Delta r_\mathrm{i}$ is flat (no considerable sinusoidal component is seen). Finally, the locking screws can be tightened to finish the concentricity adjustment.  

Fig.~\ref{fig:8}(b) and (c) present a final adjustment result, respectively for the upper and lower parts of our magnet system. The raw experimental data contain some high interference noise, and it was found due to the yoke surface quality of machining. To remove these sparks, the least squares fit is employed and the results are shown in dotted lines in Fig.~\ref{fig:8}(b) and (c). 
It can be seen from the measurement results that the peak-to-peak value of $\Delta r_\mathrm{o}-\Delta r_\mathrm{i}$ achieved is approximately 150\,$\upmu$m and 200\,$\upmu$m for the upper and lower parts, respectively.
The residual of $\Delta r_\mathrm{o}-\Delta r_\mathrm{i}$ does not exhibit a significant symmetry. The asymmetric waveforms indicate that, after the adjustment, the main reason affecting the consistency of the air gap width is the yoke deformation caused by the machining or assembly. 

\section{Experimental Results and Discussions}
\label{Sec04}

\subsection{Magnetic Profile Measurement}
An easy method for measuring the magnetic profile is to use the gradient coil~\cite{NISTmag}. A gradient coil contains two identical coils that are separated with a vertical distance $\Delta z$ and are {connected in series opposition}. Moving the gradient coil in the magnetic field with a velocity $v$ yields 
\begin{equation}
    \frac{V_1 (z)-V_2 (z)}{V_1 (z)} = \frac{B (z)-B (z-\Delta z)}{B(z)} \approx \frac{\Delta z \frac{\mathrm{d}B(z)}{\mathrm{d}z}}{B(z)},
\end{equation}
where $V_1$ and $V_2$ are the induced voltage of each coil. Hence, the profile $B(z)$ is determined as
\begin{equation}
    B(z)=\frac{\bar{B}}{\bar{V_1}\Delta z} \int(V_1(z)-V_2(z))\mathrm{d}z + B_0,
\end{equation}
where $B_0$ is chosen to ensure that $B(0)=\bar{B}$. As the variation noise of two coils is almost identical, the common-mode noise in the induced voltage can be well suppressed and a good signal-to-noise ratio is therefore achievable. 

The following is the parameter of the gradient coil used in the measurement: The main radius is 81.5\,mm, $\Delta z=6$\,mm, and the number of turns for each coil is 400. The gradient coil is connected to a moving stage (PI M-413.2DG) through three aluminum rods and the stage can drive the coil moving up and down with a constant velocity 0.5\,mm/s. The induced voltage of one coil, $V_1(z)$, and the voltage difference of two coils, $V_1(z)-V_2(z)$, are simultaneously measured by two DVMs (Keysight 3458A). 

\begin{figure}
    \centering
    \includegraphics[width=0.475\textwidth]{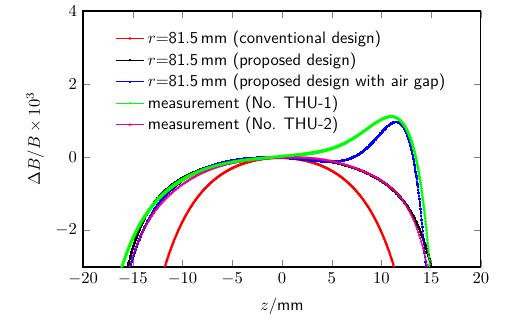}
    \caption{The measurement result of the magnetic profile. The red and black curves are theoretical profiles with and without inner-yoke-shape compensation, obtained by the FEA calculations. The green is the experimental result, in which two measurements, one up and one down, are plotted. The blue is the calculation result of an FEA model with adding a 140\,$\upmu$m  air gap on the open/close surface. The magenta profile is the experimental result when the ripple is suppressed by reducing the air gap width on the close surface.}
    \label{fig:exp_profile}
\end{figure}

Fig.~\ref{fig:exp_profile} presents the magnetic profile measurement result. Note that those profiles are relative referring to the magnetic flux density at the center point ($B=0.59\,\rm{T}$ for the NdFeB magnet, and 0.45\,T for the SmCo magnet). The red curve is a theoretical profile of a BIPM-type magnetic design without inner yoke shape compensation (parameters remain the same as the Tsinghua system). The black curve is the theoretical profile with the proposed yoke compensation ($h_\mathrm{c}=5$\,mm, $\delta_\mathrm{c}=0.4\,$mm). The green curve shows an experimental measurement result of a proposed magnet system (No. THU-1). It can be seen that the measurement agrees well with the prediction in the lower range, $z\in(-17,5)$\,mm. While in the top range, $z\in(5,15)$\,mm, the measurement result becomes higher than the theoretical value. This phenomenon results from the open/close surface: In our design, nickel coating is employed to prevent yoke rust. Although nickel has a considerable magnetic permeability, its value ($\mu_\mathrm{r}$ is typically a few hundreds) is much lower than that of the yoke. In addition, small air gaps always exist on the open/close surface. In this case, the magnetic reluctance for the upper permanent magnet flux going through the lower air gap (divided by the open/close surface) becomes increased. Then more flux goes through the air gap above the open/close surface, yielding a higher magnetic flux density. To verify the above conclusion, an FEA model with adding a tiny air gap on the open/close surface is analyzed, and the result is shown as the blue curve in Fig.~\ref{fig:exp_profile}. The FEA result agrees well with the measurement with the air gap width of 140\,$\upmu$m. The uniformity of the magnetic profile can be improved by using a thinner coating and reducing the air gap on the close surface. The magenta curve in Fig.~\ref{fig:exp_profile} presents an example (No. THU-2) where the ripple is well suppressed. 

It should be noted that the gradient coil may cause a fixed slope of the $B(z)$ measurement result when the diameters of two coils are not identical~\cite{li2022irony}. It is better to check the profile measurement result by using the $U/v$ measurement when the formal coil is integrated into the system. It is also worth mentioning that the profile measurement result is closely related to the radius of the gradient coil, $r_\mathrm{gc}$. In general, when close to the inner yoke, e.g. $r_\mathrm{gc}<r_\mathrm{a}$, the magnetic flux density at two ends of $B(z)$ profile is getting higher. While, when $r_\mathrm{gc}>r_\mathrm{a}$, the two ends of the profile drop. {It is observed that the field uniformity in the measured profile in Fig. \ref{fig:exp_profile} is lower than the predicted uniformity shown in Fig. \ref{fig:feature}(a). This discrepancy is exactly due to the use of a larger gradient coil radius ($r_\mathrm{gc}=81.5$\,mm) compared to the mean radius of the air gap ($r_\mathrm{a}=80$\,mm).}

\subsection{Alignment Test}

As mentioned in Section \ref{Subsec: Adjustment_concentricity}, the magnet must provide sufficient space for the coil to reach equilibrium during alignment. Therefore, an alignment test is essential to confirm the existence of this equilibrium position before integrating the magnet into the Kibble system.

\begin{figure}[t]
    \centering
    \includegraphics[width=0.42\textwidth]{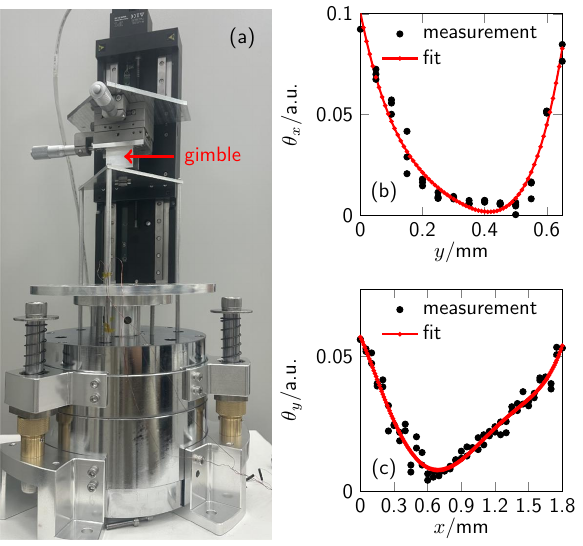}
    \caption{An example of the alignment test after the magnet is assembled. (a) shows the experimental platform. (b) and (c) present the coil oscillation amplitude as a function of the position along $y$ and $x$ directions.}
    \label{fig:10}
\end{figure}

Fig.~\ref{fig:10}(a) shows an experimental setup for the coil alignment test. The coil is mounted on an $xy$ stage through three rods and a spider. The connection between the coil suspension and the $xy$ stage is a gimble that allows the coil to horizontally move or rotate along $x$ and $y$. 
The suspension is first horizontally leveled, and before being aligned, the coil will rotate along $x$ and $y$ when an AC current is injected into the coil. Optical sensors (Mirco-Epsilon 1420) are used to measure the oscillation amplitude of the spider rotation along $x$ and $y$, i.e. $\theta_x$ and $\theta_y$. The test is to ensure valleys can be reached when moving the stage along both the $x$ and $y$ axis.  
Fig.~\ref{fig:10}(b) and (c) present an experimental result. Note that the frequency of the current excitation should make the coil rotation be sensitive to $xy$ movement. In this test, the frequency of the current through the coil is set to 1\,Hz. It can be seen from the measurement that both $\theta_x$ and $\theta_y$ can reach the minimum and therefore the test coil can reach an equilibrium where a good coil alignment is achievable.  

\subsection{Temperature Characteristic}
\label{ssec:TempC}
    The thermal dependence is an important parameter for Kibble balance magnetic circuits, especially crucial for tabletop systems~\cite{li2022}. Typically, the NdFeB temperature coefficient is approximately $-1\times10^{-3}$\,/K and the SmCo magnet is about $-3\times10^{-4}$\,/K. Although most Kibble balances by far choose the SmCo magnet system due to a low-temperature coefficient, it is still very attractive to use the NdFeB magnet as the magnetic field created here is over 30\% stronger than that of the SmCo magnet. The magnetic field change due to the environmental temperature drift, in principle, can be well eliminated by ABA or ABBA measurements. Instead, an important concern is the systematic effects caused by the noncontinuous coil ohmic heating, which can not be removed in a conventional Kibble measurement scheme, see \cite{li2022}. A stronger magnetic field, in this case, can lead to a reduction of the coil ohmic heating (if the same wiring is adopted) and the current effect, which may compensate for or even lower the thermal-related bias. 

\begin{figure}[tp!]
    \centering
    \includegraphics[width=0.475\textwidth]{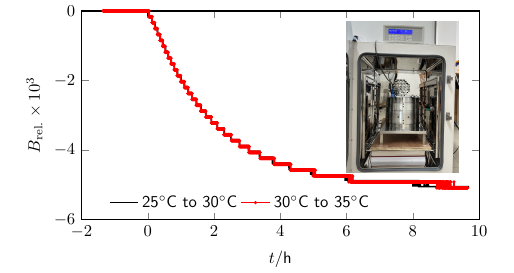}
    \caption{Measurement results of temperature characteristics for the constructed magnet system (NdFeB version). }
    \label{fig:temp}
    \end{figure}

    The temperature coefficient of the magnet system can be measured in several different ways, and the most accurate result comes from the measurement of $Bl$ in the final stage along with the temperature variation. Without a surprise, the measurement result should agree more or less with these values given by the permanent magnet manufacturers. It is believed the thermal time constant, which is independent of the magnet type (NdFeB or SmCo) and mainly depends on the size and material of the segments forming the magnet system, is a more important parameter. Here a measurement of the thermal time constant and temperature coefficient of an NdFeB system is presented. As shown in the subplot of Fig.~\ref{fig:temp}, the magnet system is placed in a temperature-controlled oven. Its temperature control accuracy is $\pm 0.2\,\rm{^\circ C}$. Two sets of temperature ramping, respectively $25\,\rm{^\circ C}$ to $30\,\rm{^\circ C}$, and $30\,\rm{^\circ C}$ to $35\,\rm{^\circ C}$, are carried out. {A Gauss meter (Lakeshore} Model 425) is used to measure the air-gap magnetic flux density change. 

    The experimental results are shown in Fig.~\ref{fig:temp}. The oven can reach the targeted temperature ($30\,\rm{^\circ C}$ and $35\,\rm{^\circ C}$) within a few minutes, while the reduction of the magnet flux density lasts about 10 hours to reach stability. The measurement results show that the relative change of the magnetic flux density for the two tests is almost identical. An exponential fit of the measurement result yields a time constant of 2\,h and the temperature coefficient obtained is about $-1.1\times 10^{-3}$\,/K. 
    
\subsection{$Bl$ and the Current Effect}

    With the magnet system finalized, the $Bl$ value and the current effect \cite{li17} of the experiment can be evaluated. Here we take the Tsinghua system as an example: For the NdFeB version, the average magnetic flux density at the air gap is about 0.59\,T. The inner and outer radii of the coil wiring section are 75.5\,mm and 86\,mm, and its height is 15\,mm. The number of turns for each coil for the bifilar coil is 1360 (0.2\,mm wire gauge) and the resistance of each coil is measured at approximately 395\,$\Omega$. In this configuration, the $Bl$ value given by each coil is over 400\,Tm. If a conventional two-mode, two-phase measurement scheme is chosen, the $Bl$ is doubled and is over 800\,Tm. A $Bl$ value in the range of a few hundred Tm allows the experimental to achieve the lowest uncertainty for a kilogram mass realization~\cite{schlamminger2013design,li2022irony}. 

    The additional magnetic field generated by the coil current reshapes the magnetic field distribution in the air gap and hence affects the weighing and velocity measurements (the velocity measurement is affected only when the one-mode scheme is adopted). There are a few approaches to determine the coil current effect, detailed in \cite{li17}. Here a direct measurement of the coil inductance is used. The coil is set to a fixed position in the air gap of the magnet, $z$. A 200\,$\Omega$ resistor ($R_\mathrm{s}$) is in series with the coil and is excited by a low-frequency current. The voltages of $R_\mathrm{s}$ and the coil are simultaneously sampled by two DVMs (Keysight 3458A). The amplitude and phase of the two measurements can be extracted, denoted as $V_R$, $\phi_R$ and $V_L$, $\phi_L$ respectively for the resistor and the coil. The following equations allow to solve the coil resistance and coil inductance, i.e. 
    \begin{eqnarray}
        \frac{R_\mathrm{s}^2}{R_\mathrm{s}^2+(\omega L)^2}=\left(\frac{V_R}{V_L}\right)^2,\\
        \frac{\omega L}{R_L}=\tan(\phi_L-\phi_R),
    \end{eqnarray}
    where $\omega=2\pi f$ is the angular velocity of the current applied. Measuring the $L$ value at different coil positions can yield the $L(z)$ curve. 

    \begin{figure}
        \centering
        \includegraphics[width=0.475\textwidth]{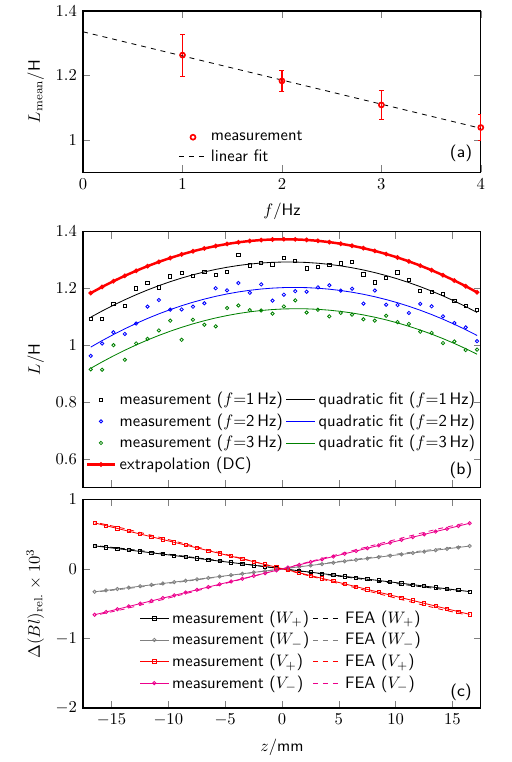}
        \caption{{(a) shows an estimation of mean coil inductance within 35\,mm as a function of the measurement frequency.} (b) presents the measurement result of the coil inductance, $L(z)$, at different frequencies ($f=1,2,3$\,Hz). The red curve is a linear extrapolation of the quadratic fit of the measurements, representing the $L(z)$ value at DC.  (c) {compares} the relative $Bl$ change due to the coil currents, respectively obtained by the $L(z)$ measurement and the FEA calculation. $W_+$, $W_-$, $V_+$ and $V_-$ denote respectively weighing with positive current, weighing with negative current, velocity measurement with positive current, and velocity measurement with negative current \cite{BIPM,li2022design}.}
        \label{fig:12}
    \end{figure}
    
    Note that the $L$ value required here should be a DC value. To remove the frequency dependence, the measurement was carried out with different current frequencies, $f=1,2,3,4$\,Hz. {Fig. \ref{fig:12}(a) illustrates the mean inductance, $L_\mathrm{mean}$, within a range of $\pm 17.5\,\rm{mm}$  as a function of the current frequency $f$. The measurement results indicate that the inductance varies linearly with $f$ for frequencies up to 4\,Hz. This linear relationship allows for the extrapolation of the inductance's DC value. Fig. \ref{fig:12}(b) presents detailed measurements of $L(z)$ at frequencies of 1, 2, and 3\,Hz.} A quadratic fit is applied to these measurements, and further, a linear extrapolation is used to determine the $L(z)$ curve, see the red curve in Fig.~\ref{fig:12} (b). The quadratic fit gives: $L(z)=-650.3\,\mathrm{H/m}^2z^2+1.3725\,\mathrm{H}$. The constant term denotes the maximum $L$ value where the coil is at the symmetrical center, $z=0$\,mm. Using $\partial L/\partial z$, the magnetic profile change due to the coil current can be calculated, presented in Fig.~\ref{fig:12}(c). As a comparison, the FEA result, obtained by $F/I$ calculation, is also shown in Fig.~\ref{fig:12}(c). It can be seen that the measurement result agrees well with the FEA calculation. 

\section{Conclusion}
\label{Sec05}
How to develop a compact magnet system is an important task for tabletop Kibble balance experiments. A direct volume reduction of the conventional BIPM-type magnet may lead to a considerable increase in measurement uncertainty, and hence necessary optimization of the magnet system is required to balance the measurement uncertainty and the overall size. Hereby, we present the design and realization of a compact magnet system for the Tsinghua tabletop Kibble balance. Some noticeable features are achieved:

\begin{itemize}
    \item 
    The magnet presented is divided into upper and lower parts by an optimal open/close surface. When closed (separation distance $d=0$\,mm), the two parts are tightened by an attractive magnetic force. When opened by a few millimeters, the magnetic force becomes repulsive, allowing the upper part to levitate robustly. This easy open/close operation facilitates the maintenance of the magnet system within the experiment. Additionally, the opening and closing forces are reduced to tens of kilograms, compared to the kN-level attraction force of conventional designs, enhancing operational convenience.
    
    \item 
    The magnet system is optimized to achieve a sufficient $Bl$ by extending the range of the one-dimensional radial magnetic field. The inner yoke shape is modified to improve magnetic field uniformity in the $z$ direction, and circumferential uniformity is enhanced through optimization of the yoke material and adjustment of the concentricity between the inner and outer yokes. As a result, the Tsinghua system achieves a $Bl$ over 400\,Tm for a bifilar coil (800\,Tm for a single coil) using 0.2\,mm wire gauge.
    
    \item Following the open-hardware idea, details of the magnet assembly and adjustments are provided. A precision assembly of the magnet system was achieved by using very simple tools, and some readily available methods are presented for fine adjusting and characterizing the performance of the realized magnet system. This provides useful references for building such magnet systems.
\end{itemize}

In subsequent sections, the realized magnet system will be integrated into the Tsinghua tabletop Kibble balance measurement, with careful evaluation of related systematics. We believe that the proposed magnet system meets the high-accuracy requirements for kilogram-level mass realizations in Kibble balance experiments.

\section*{Acknowledgement}
The authors would like to thank our mechanical engineer Mr. Zhenyu Zhang for his help on the magnet design and assembly. Shisong Li would like to thank colleagues from the NIST Kibble balance group and the BIPM Kibble balance group for valuable discussions. Yongchao Ma would like to thank Mr. Jian Liu and Mrs. Zhilan Huang for their help during the assembly of the magnet system. 


\end{document}